\documentstyle[epsf,prl,aps]{revtex}

\title{The self-similarity properties of natural images resemble those 
of turbulent flows}

\author{
Antonio Turiel, Germ\'an Mato, N\'estor Parga
\\
Departamento de F\'{\i}sica Te\'orica \\
Universidad Aut\'onoma de Madrid \\
Cantoblanco, 28049 Madrid, Spain \\
and \\
Jean-Pierre Nadal\\
Laboratoire de Physique Statistique de l'E.N.S.\thanks{Laboratoire 
associ\'e au C.N.R.S. (U.R.A. 1306), \`a l'ENS, et aux Universit\'es 
Paris VI et Paris VII.}\\
Ecole Normale Sup\'erieure\\
24, rue Lhomond, F-75231 Paris Cedex 05, France}

\begin{document}

\maketitle

\begin{abstract}
We show that the statistics of an edge type variable in natural images exhibits
self-similarity properties which resemble those of local energy dissipation
in turbulent flows.
Our results show that extended self-similarity 
remarkably holds for the statistics 
of the local edge variance,
and that the very same models can be used to predict all the associated
exponents. These results suggest
to use natural images as a laboratory for testing scaling models
of interest for the statistical description of turbulent flows.
The properties we have exhibited are relevant for the modeling 
of the early visual system~:
they should be included in models designed for the prediction of receptive fields.
\end{abstract}

\vspace{1cm}
PACS numbers 42.30.Yc, 47.27.Gs, 05.70.Jk, 42.66.Lc

{\it Physical Review Letters}, {\bf 80}(5), 1098-1101. (1998)

\twocolumn

The existence of self-similarity (SS) is well known in both
natural images 
\cite{Field} 
and fully developed turbulence 
\cite{K41}.
Quite recently, there has been an increase of interest in both fields.
In turbulent flows, the notion of ``extended self-similarity'' (ESS)
\cite{ESS,ESSbig,ESS-exp}
has been introduced, and several models proposed predicting
correctly the relevant SS exponents from only one or two parameters
\cite{SheLeveque,Castaing}.
Our main motivation for studying the statistics of natural images
is its relevance for the modeling of the early visual system. In particular,
the epigenetic development could lead to the adaptation of
visual processing to the statistical regularities in the visual scenes 
\cite{barlow61,laughlin,hateren,atick_rev,OF_nature,Baddeley}.
Most of these predictions on the development of receptive fields
have been obtained using a gaussian description 
of the environment contrast statistics. 
However  non Gaussian properties like the ones 
found by  \cite{RudermanBialek,Ruderman}
could be important.
To gain further insight into 
non Gaussian aspects of natural scenes 
we investigate whether they exhibit  the rich structure found 
in turbulent flows.

Scaling properties of natural images
have been studied by several authors. 
They have found \cite{Carlson,burton_etal,Field}
that the power spectrum of luminosity contrast follows a power law  
of the form $S(f) \propto \frac{1}{|f|^{2-\eta}}$, although the 
value of $\eta$ can have rather large fluctuations 
\cite{Tollhurst}. The magnitude of these fluctuations 
depends on the diversity of the images in the data set. 
A more detailed - although different - analysis of the scaling properties  
of image contrast  was done by \cite{RudermanBialek,Ruderman}
who also noted analogies with the statistics of turbulent flows. 
Additional luminosity analysis was also done by D. Ruderman \cite{private},
providing some evidence of multiscaling behaviour.
There is however no model to explain the intriguing scaling behavior observed. 

On the contrary in turbulent fluids
the unpredictable character of signals 
has led to a large amount of effort in order to develop
statistical models (see e.g. \cite{Frisch}).
Qualitative and quantitative theories of the statistical properties of
fully developed turbulence elaborate on the original argument of 
Kolmogorov \cite{K41}. The cascade of energy from one scale to another is 
described in terms of local energy dissipation 
 per unit mass within a box of linear size $r$. 
This quantity,  $\epsilon_r$, is given by:
\begin{equation}
\epsilon_r({\bf x}) \propto \int_{|{\bf x}-{\bf x}'|<r} \; d{\bf x}' \sum_{ij} [\partial_i v_j({\bf x}') +\partial_j v_i({\bf x}')]^2
\label{eq:epsilon}
\end{equation}
where $v_i({\bf x})$ is the $i$th component of the velocity at point ${\bf x}$.
{\it Self-Similarity} (SS) will hold, if for some range of scales $r$
one finds the scaling relation:
\begin{equation}
< \epsilon_{r}^p > \; \propto \; r^{\tau_p}
\label{tau}
\end{equation}
where $< \epsilon_{r}^p >$ denotes the $p$th moment of the energy dissipation, 
that is the
average of $[\epsilon_r({\bf x})]^p$ over all possible values of ${\bf x}$.
In fluid dynamics this property holds in the so-called ``inertial
range" \cite{Frisch}. 
A more general scaling relation, called {\it Extended Self-Similarity}
(ESS) has been found to be valid in a much larger scale domain,
even if the inertial range does not exist \cite{ESS,ESSbig}.
This scaling can be defined by:
\begin{equation}
<\epsilon^p_{r}> \; \propto \; <\epsilon^q_{r}>^{\rho(p,q)}
\end{equation}
where $\rho(p,q)$ is the ESS exponent of the $pth$ moment with respect 
to the $qth$ moment. 
 Let us notice that if SS holds then $\tau_p = \tau_q \rho(p,q)$.
In the following we will refer all the moments to $<\epsilon^2_{r}>$.

The basic field in turbulence is the velocity from which one defines
the local energy dissipation.
The largest contributions to $\epsilon_{r}$
come from abrupt changes in velocities.
For images, the basic field is the contrast $c({\bf x})$, 
that we define as the difference between the luminosity and its 
average.
A natural candidate for a variable 
analog to the local energy dissipation is 
a quantity which takes its largest contributions from the places
where large changes in contrast occur.
This is precisely a measure of the existence of  
edges below the  scale under consideration. 
Edges are indeed well known to be very important 
in characterizing images \cite{Marr}.
A recent numerical analysis suggests that natural images
are composed of statistically independent
edges \cite{BellSej-edges}.

We choose to study two variables, defined at  position 
${\bf x}=(x_1,x_2)$ and at scale $r$.
The variable $\epsilon_{h,r}({\bf x})$ takes contributions from 
edges transverse to a {\it horizontal} segment of size $r$,
that is from the derivative of the contrast along the horizontal direction:
\begin{equation}
\epsilon_{h,r}({\bf x}) = \frac{1}{r} \int_{x_1}^{x_1+r}
\left. \left( \frac{\partial c({\bf x}')}{\partial y}\right)^2 \right|_{{\bf x}'=\{y,x_2\}} dy
\label{eq:edge-horiz}
\end{equation}
A vertical variable $\epsilon_{v,r}({\bf x})$ is 
defined similarly from an integration
over the vertical direction. From here we see that 
$\epsilon_{l,r}({\bf x})$ ( $l=h,v$ ) is the {\em local linear 
edge variance} along the direction $l$ at scale $r$.

We have analyzed the scaling properties of the 
local linear edge variances 
in a set of 45 images taken in the wood, of 
$256 \times 256$ pixels each (the images have been kindly
provided to us by D. Ruderman;
see \cite{Ruderman} for technical details concerning these images).
On these data one can explore scales up to $r\sim 64$ pixels.

\begin{figure}[thb]
\hspace*{1.6cm}
$\ln<\epsilon_{r}^{2}>$
\hspace*{3.5cm}$a$
\hspace*{5cm}$b$ 
\begin{center}
\makebox[2cm]{$\ln<\epsilon_{r}^{3}>$}
\leavevmode
\epsfysize=4cm
\epsfbox[50 61 388 291]{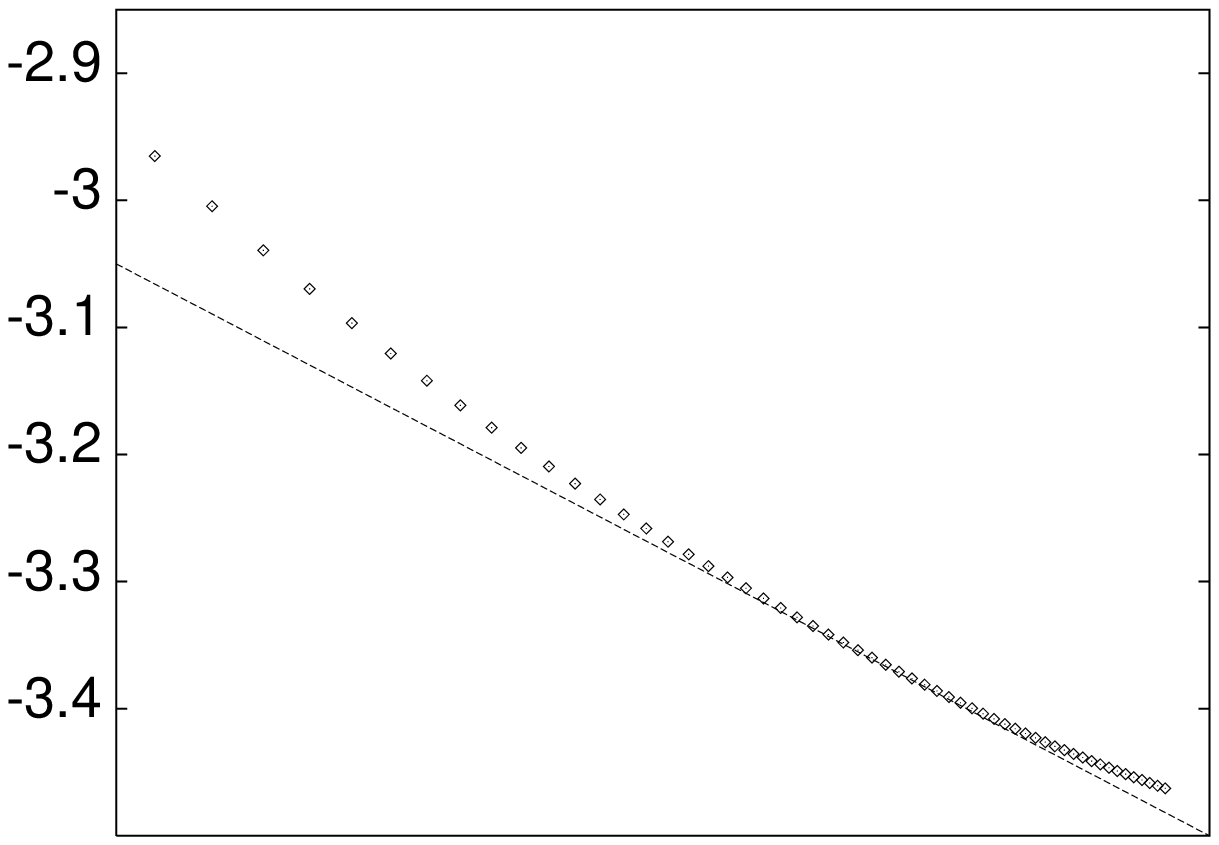}
\epsfysize=4cm
\epsfbox[72 61 410 291]{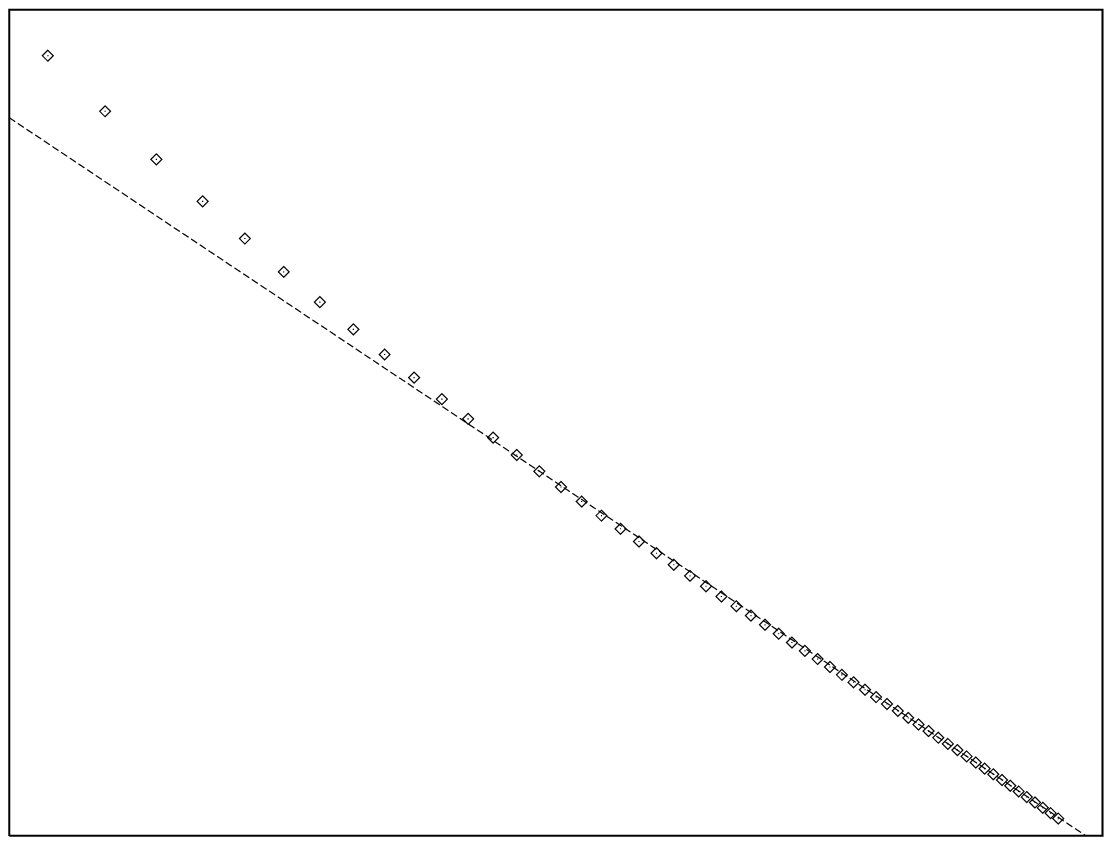}
\\
\makebox[2cm]{$\ln<\epsilon_{r}^{5}>$}
\leavevmode
\epsfysize=4cm
\epsfbox[50 61 388 291]{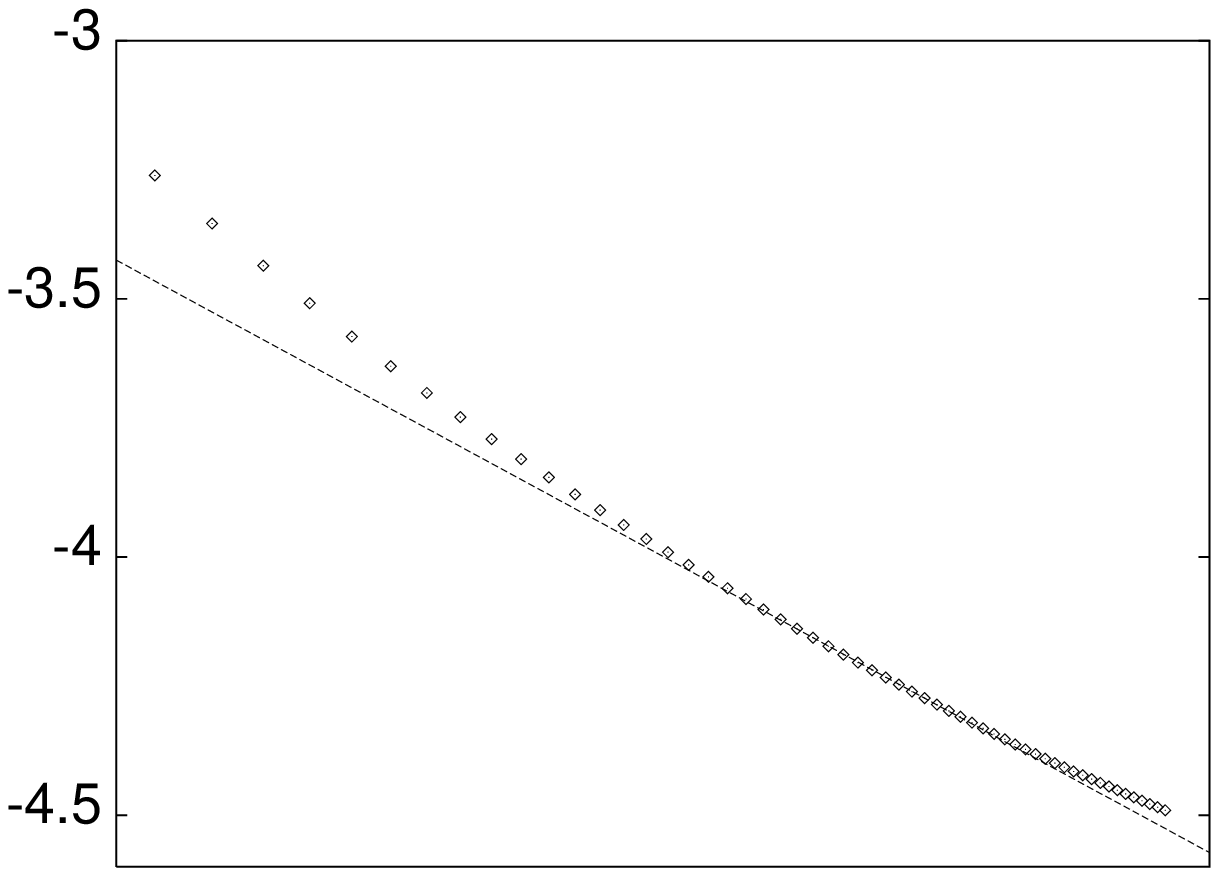}
\epsfysize=4cm
\epsfbox[72 61 410 291]{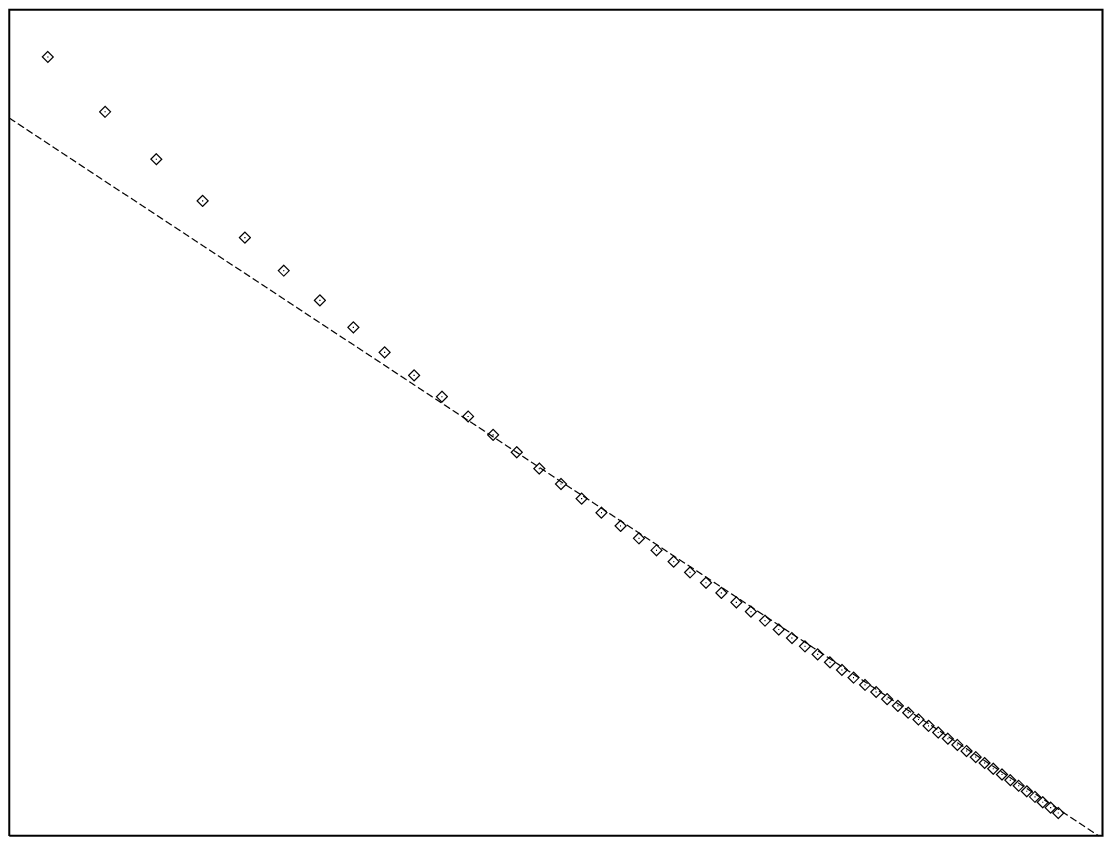}
\\
\leavevmode
\makebox[2cm]{}
\epsfysize=4cm
\epsfbox[50 61 388 291]{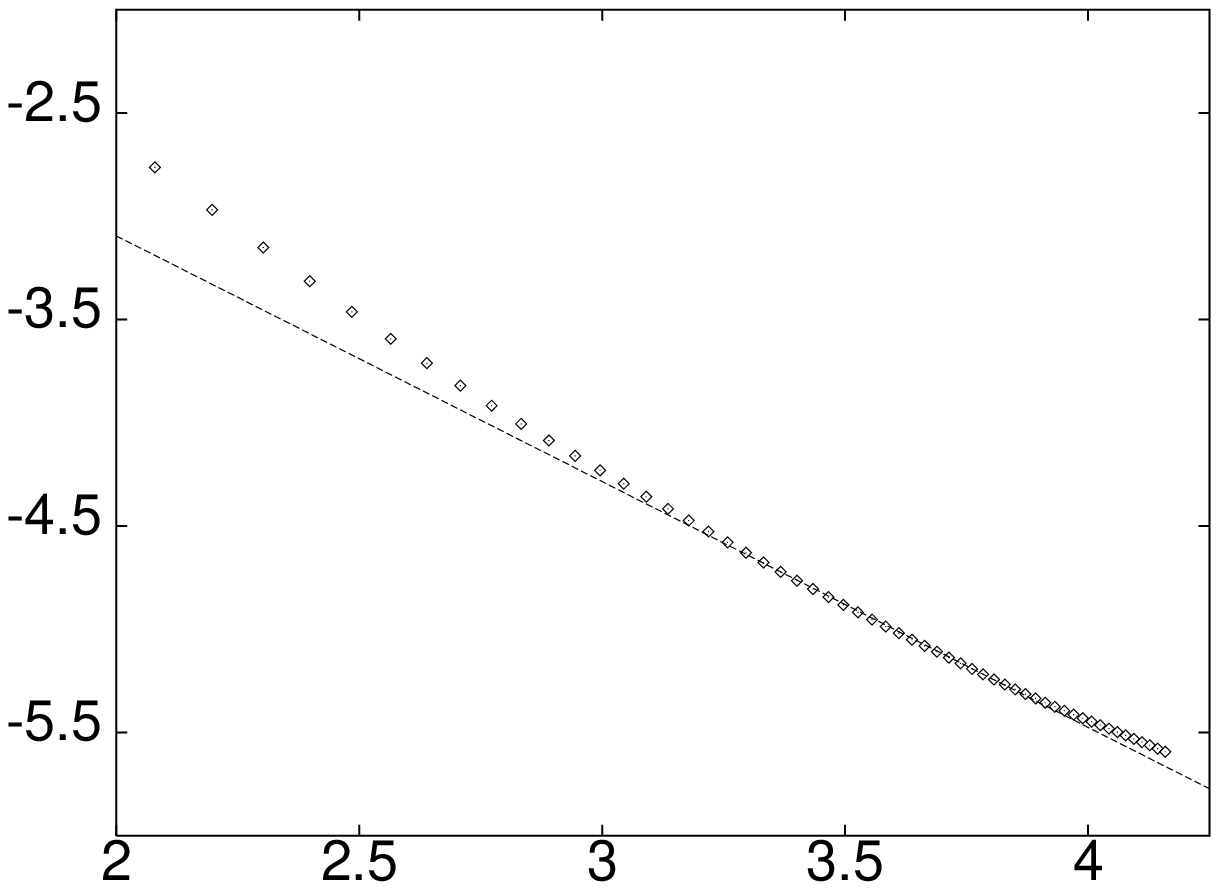}
\epsfysize=4cm
\epsfbox[72 61 410 291]{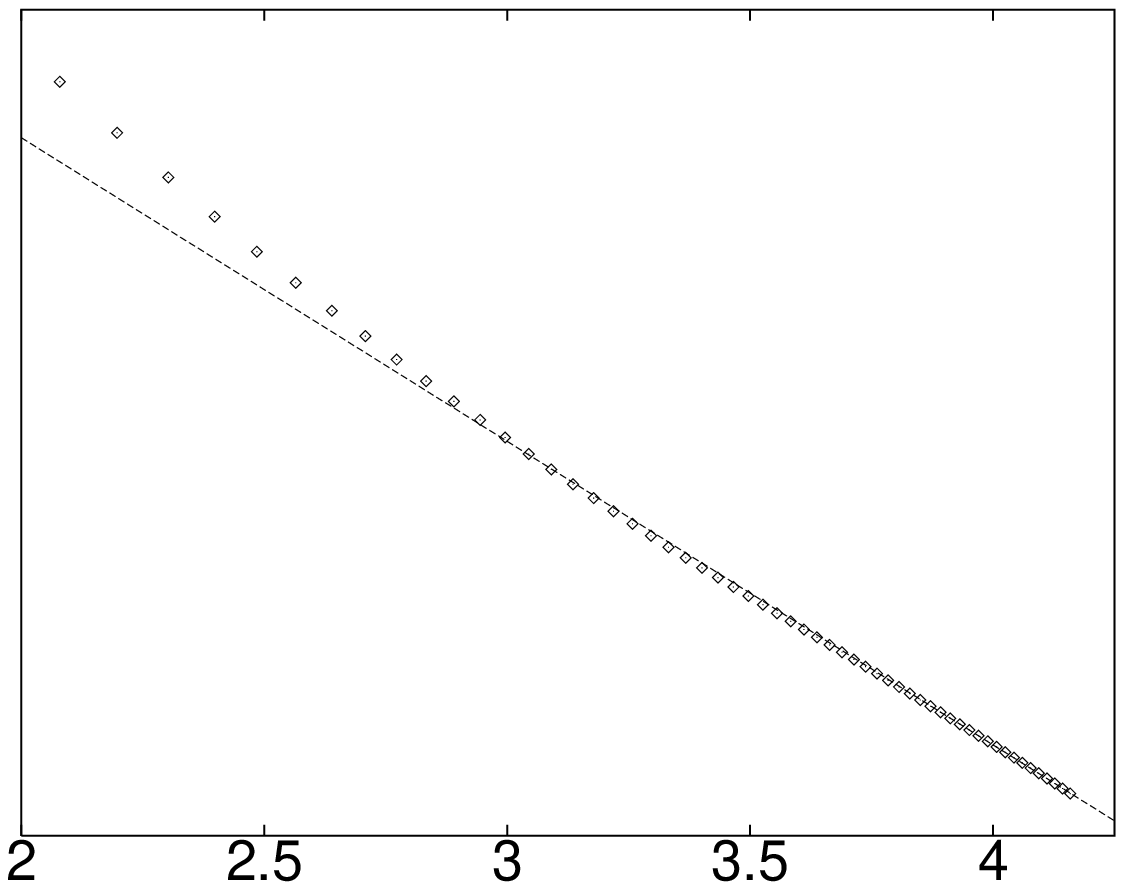}
\\
\vspace*{.25cm}
\hspace*{3.5cm}$\ln r$
\hspace*{5cm}$\ln r$
\vspace*{.5cm}
\caption{ Test of SS. We plot
$\protect\ln <
\protect\epsilon_{l,r}^{p} >$ vs.
$\protect\ln r$ for $p = 2, 3$ and $5$ and for $r$ from 
$8$ to $64$ pixels. 
a) horizontal direction, $l=h$.
b) vertical direction, $l=v$. 
The relative error is uniform and about $8 \%$.
The value of the SS exponents $\protect\tau_p$ extracted from the large 
r behavior are: $\protect\tau_{h,2} = -0.20\protect\pm 0.01 $, 
$\protect\tau_{h,3} = -0.51\protect\pm 0.02 $,  and $\protect\tau_{h,5} 
= -1.19\protect\pm 0.06$ for the horizontal direction and 
$\protect\tau_{v,2} = -0.26\protect\pm 0.04 $, $\protect\tau_{v,3} = 
-0.62\protect\pm 0.03$ and $\protect\tau_{v,5} = -1.47\protect\pm 0.06 $ for 
the vertical direction. The represented solid lines have the slope given
by these exponents. This linear behavior does not hold at small 
r. A numerical analysis indicates that it is a finite resolution effect 
although it could be masking a different, small $r$ regime. There is also an 
upper bound that has prevented us from going beyond 
$r \protect\sim 64$.}
\label{SS.gr}
\end{center}
\end{figure}

\begin{figure}[thb]
\hspace*{1.6cm}$\ln<\epsilon_{r}^{3}>$
\hspace*{3.5cm}$a$
\hspace*{5cm}$b$ 
\begin{center}
\makebox[2cm]{$\ln<\epsilon_{r}^{5}>$}
\leavevmode
\epsfysize=4cm
\epsfbox[50 61 388 292]{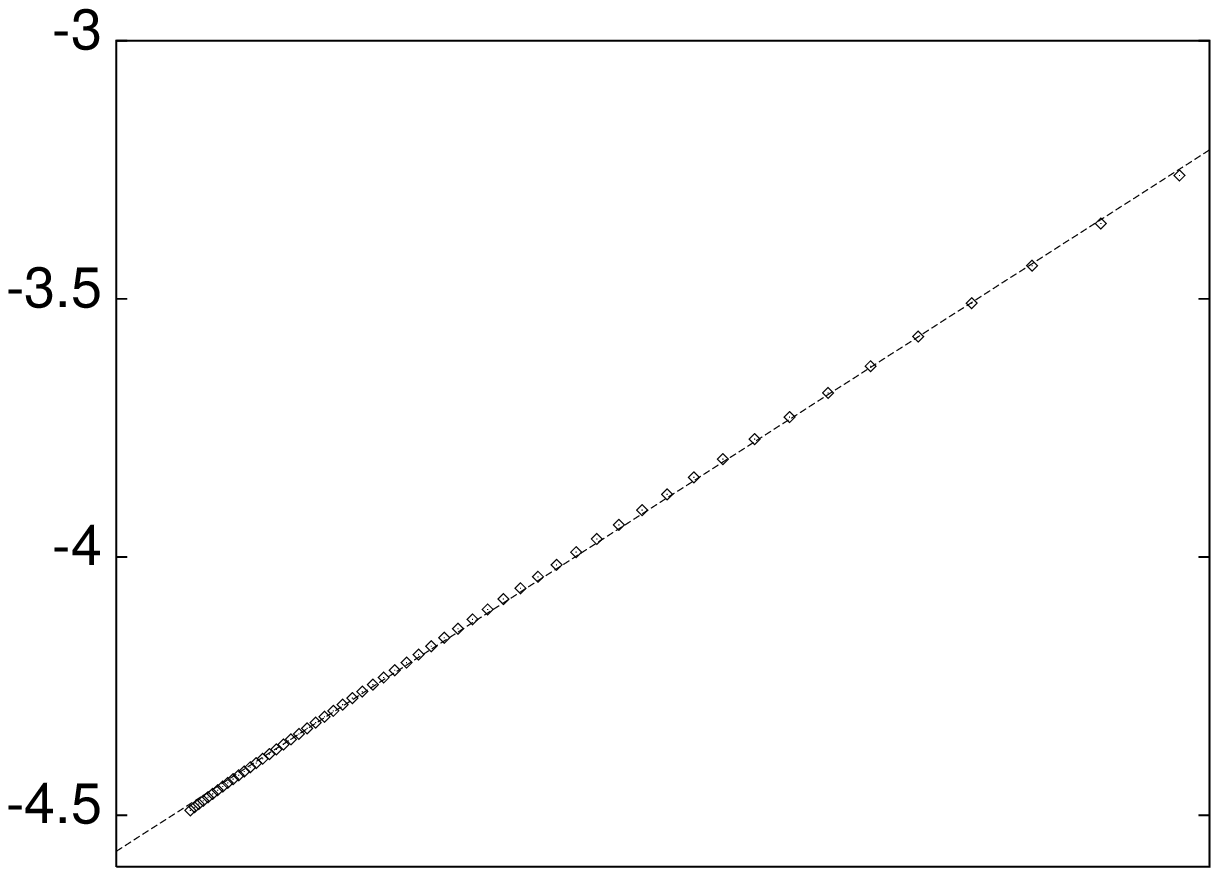}
\epsfysize=4cm
\epsfbox[72 61 410 292]{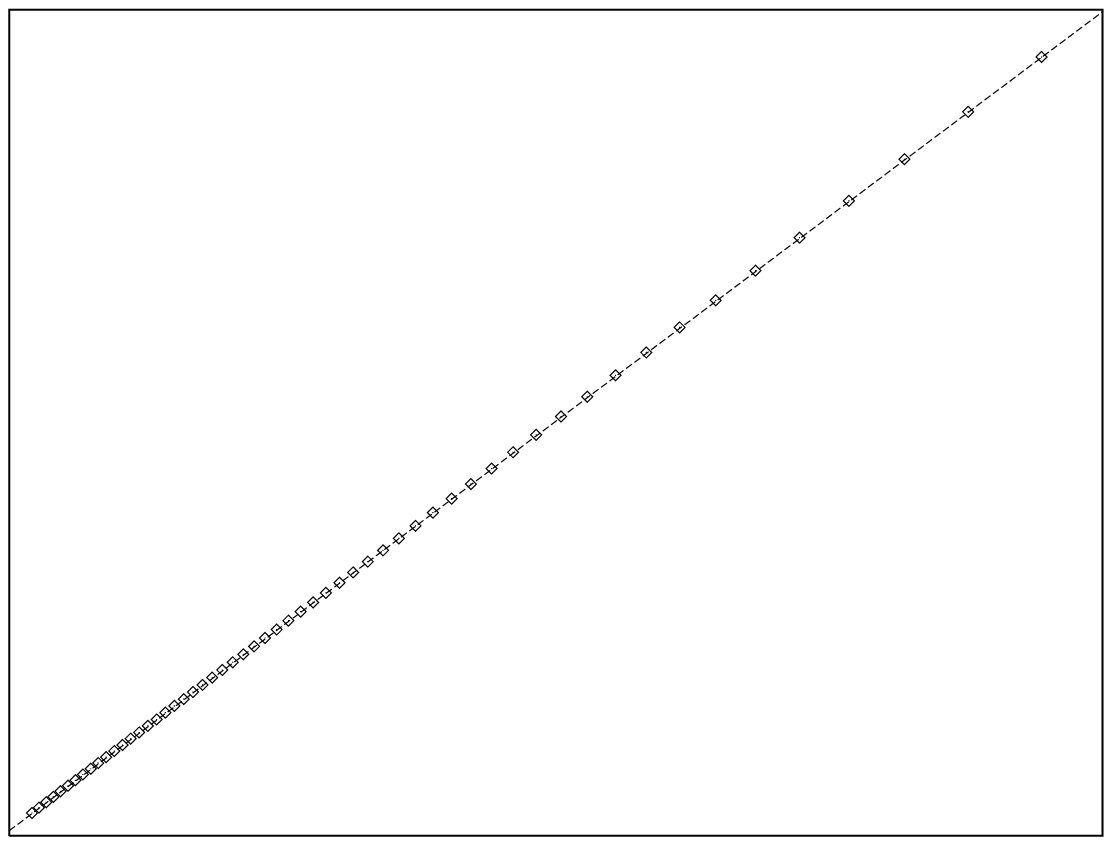}
\\
\makebox[2cm]{$\ln<\epsilon_{r}^{10}>$}
\leavevmode
\epsfysize=4cm
\epsfbox[50 60 388 292]{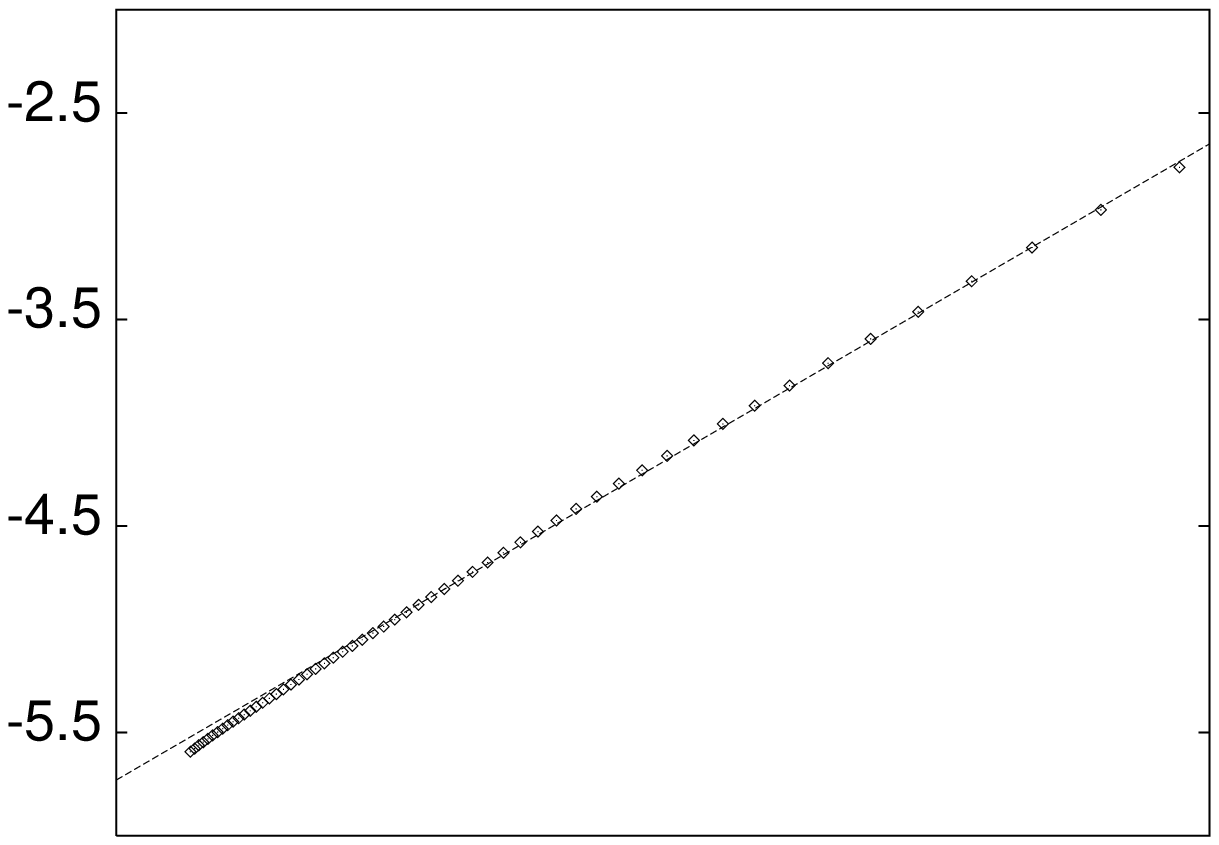}
\epsfysize=4cm
\epsfbox[72 60 410 292]{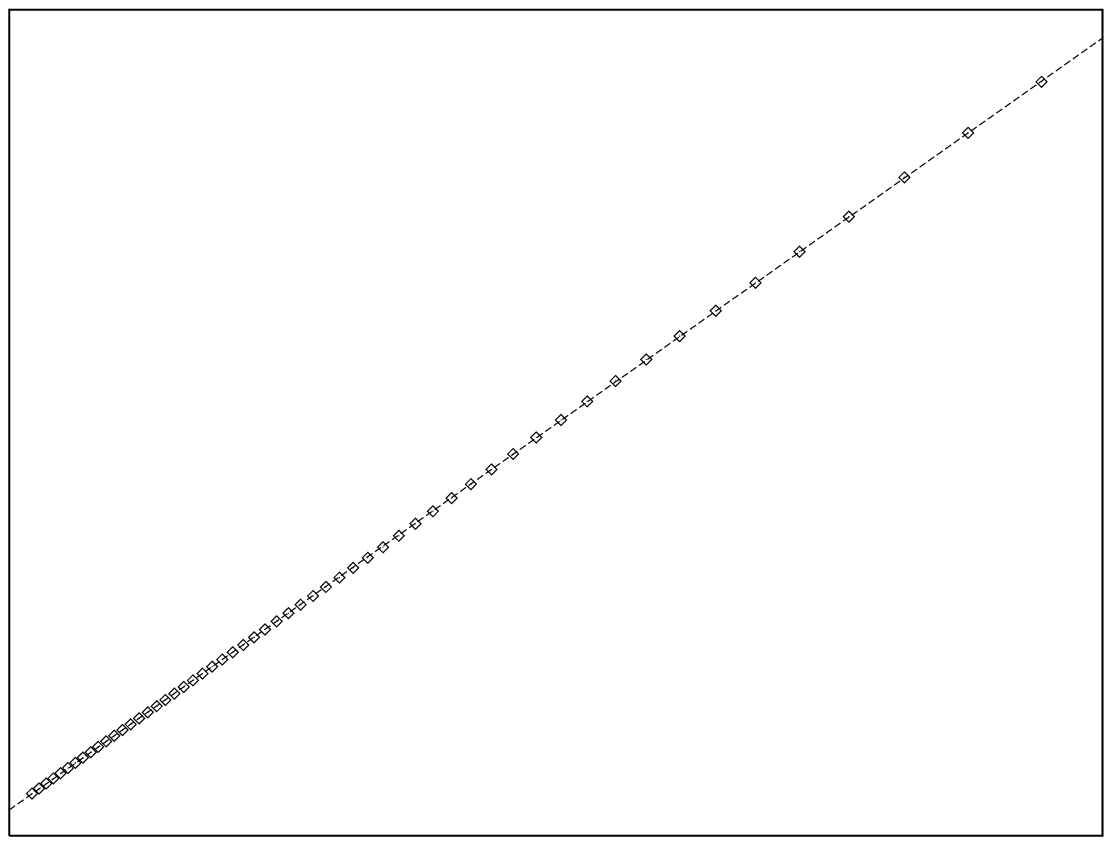}
\\
\leavevmode
\makebox[2cm]{}
\epsfysize=4cm
\epsfbox[50 60 388 292]{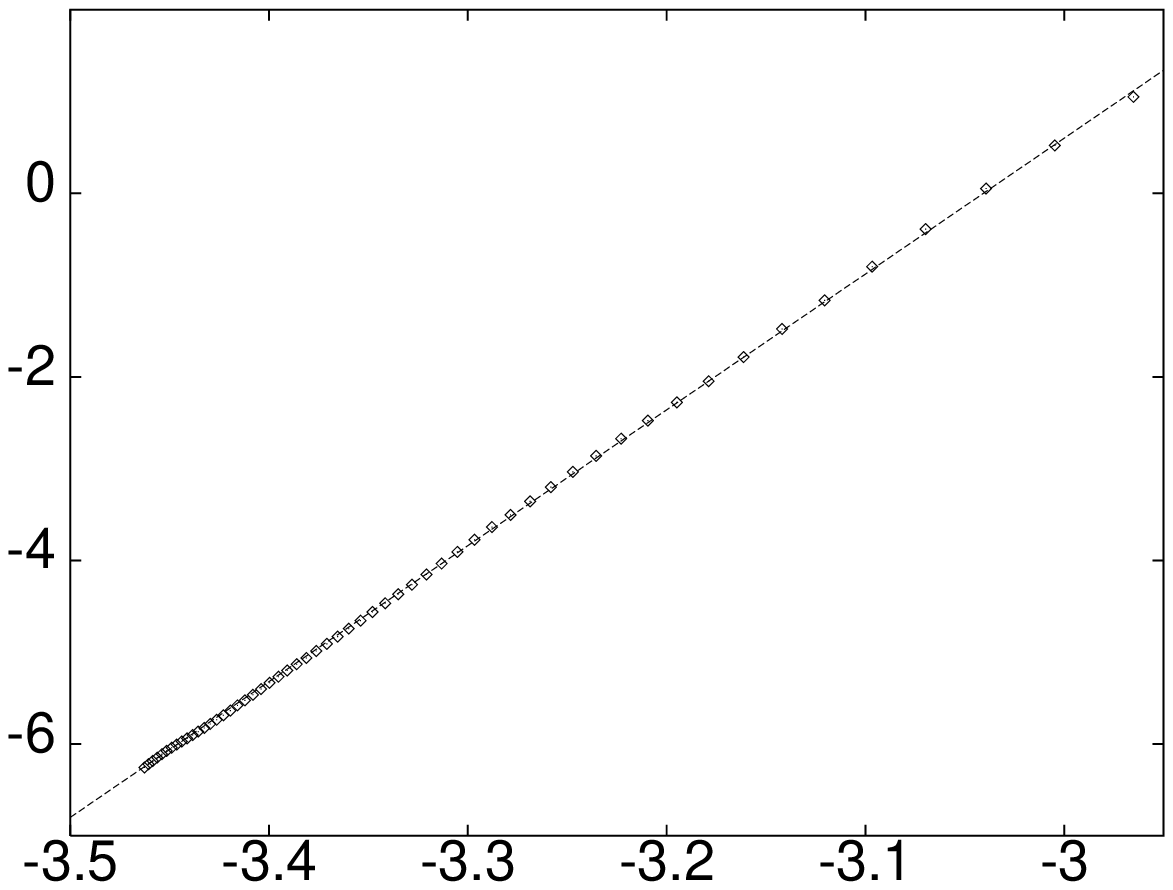}
\epsfysize=4cm
\epsfbox[72 60 410 292]{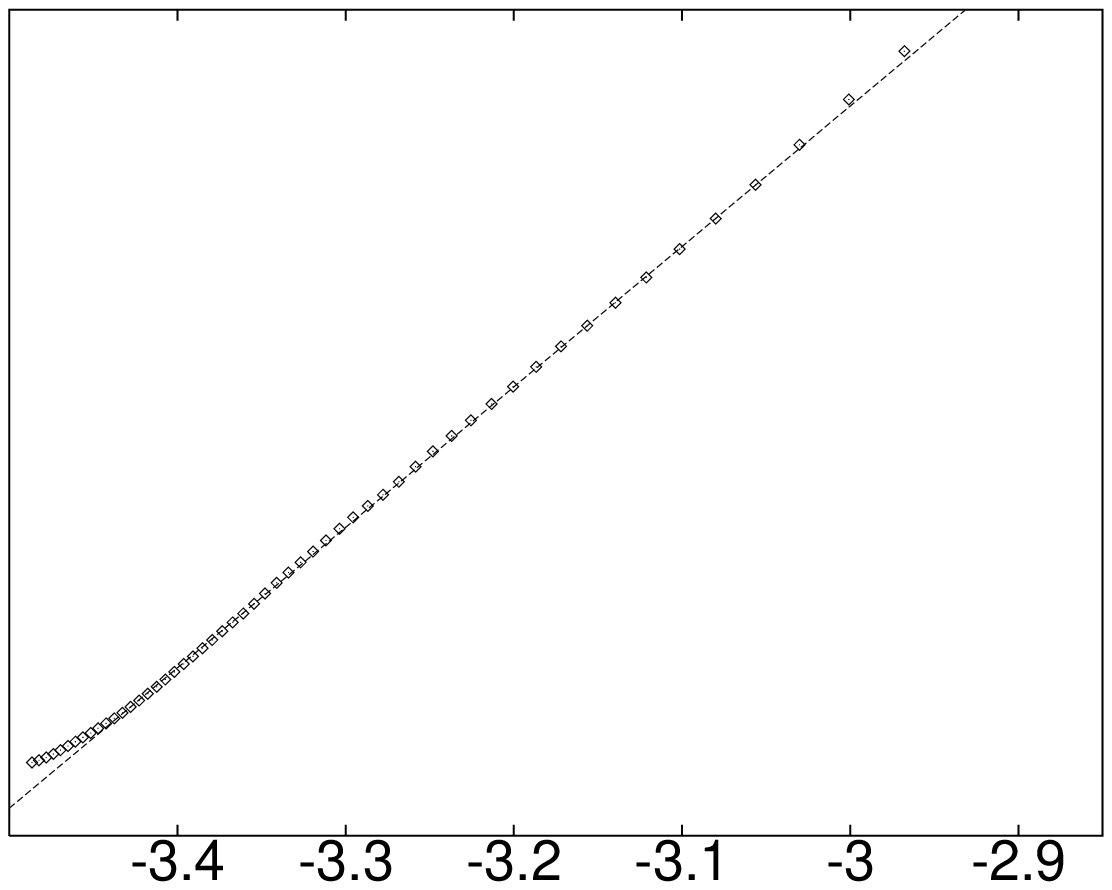}
\\
\vspace*{.25cm}
\hspace{2.5cm}$\ln<\epsilon_{r}^{2}>$
\hspace{4cm}$\ln<\epsilon_{r}^{2}>$
\vspace*{.5cm}
\caption{ Test of ESS. We plot $\protect\ln < \protect\epsilon_{l,r}^{p} >$ vs.
$\protect\ln < \protect\epsilon_{l,r}^2 >$
for p=3, 5 and 10. Data corresponds to scales  from $r=8$  to $r=64$ pixels. 
The effect of finite size effects can again be observed for $r$ close to 
$64$ pixels.
a) horizontal direction, $l=h$.
b) vertical direction, $l=v$.
The represented solid lines have the slope given
by the calculated exponents $\protect\rho(p,2)$.}
\label{ESS.gr}
\end{center}
\end{figure}

\begin{figure}[htb]
$\rho (p,2)$
\hspace*{3.75cm}$a$
\hspace*{7.25cm}$b$
\begin{center}
\leavevmode
\epsfysize=6cm
\epsfbox[50 50 388 302]{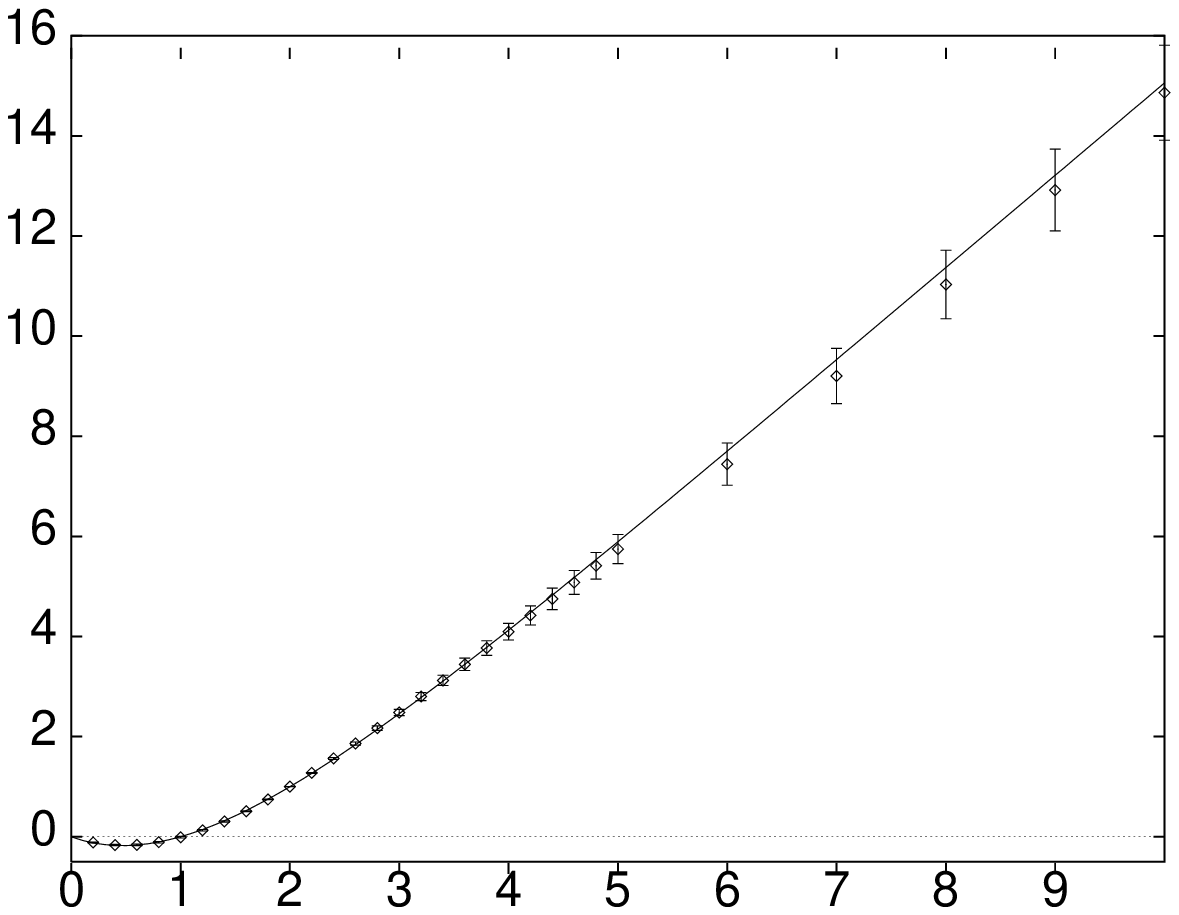}
\epsfysize=6cm
\epsfbox[72 50 410 302]{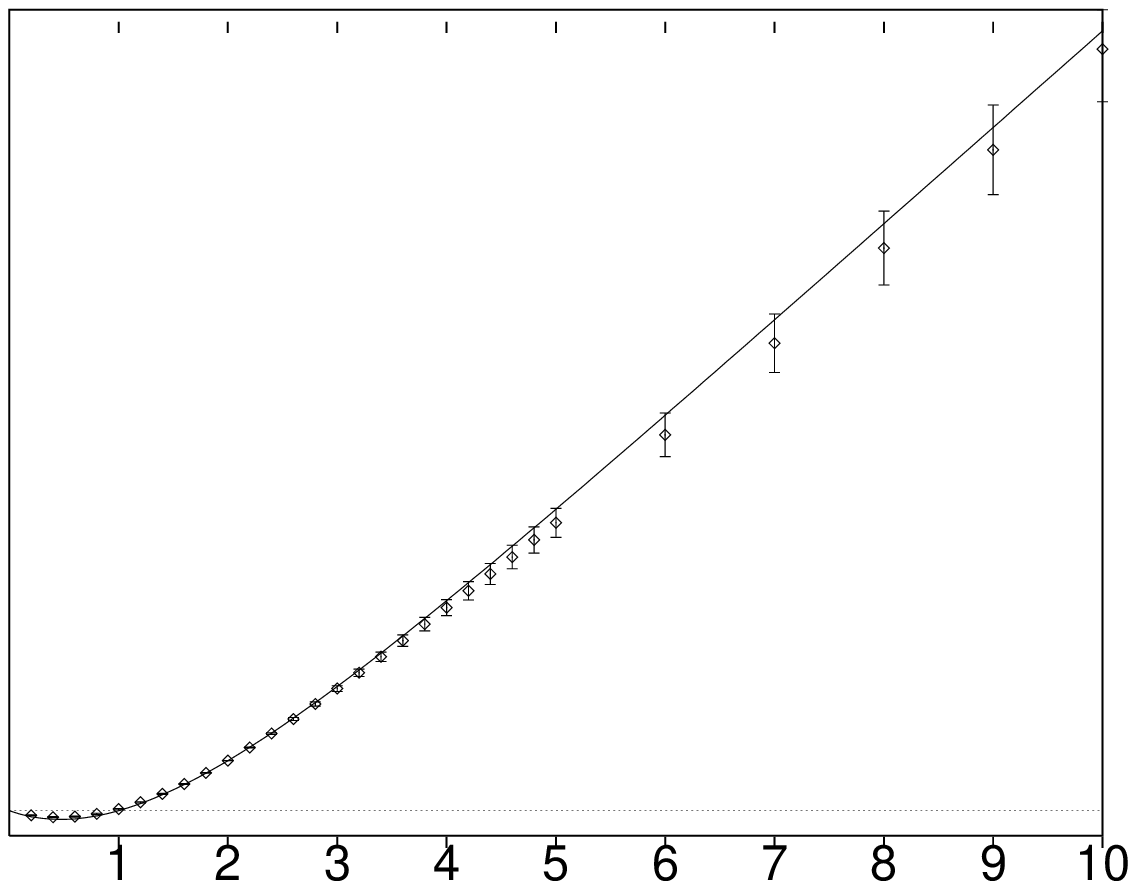}
\vspace*{.1cm}
\hspace*{.5cm}$p$\hspace*{7.5cm}$p$
\vspace*{.5cm}
\caption{ ESS exponents $\protect\rho(p,2)$, for the vertical and 
horizontal variables. 
Each value of $\protect\rho_l(p,2)$ was
obtained by a linear regression of  
$\protect\ln < \protect\epsilon_{l,r}^{p} >$ vs $\protect\ln <
\protect\epsilon_{l,r}^2 >$ for distances
$r$ between 8 and 64 (l = v, h).
a) horizontal direction, $\protect\rho_h(p,2)$.
b) vertical direction, $\protect\rho_v(p,2)$.
The solid line represents the fit with the SL model. 
The best fit  is obtained with $\protect\beta_v \sim 
 \protect\beta_h \sim 0.50$. 
The  error bars $b_p$ have been estimated by  
 dividing the 45 images in 9 groups, evaluating  
$\protect\rho_l(p,2)$ for each 
of them and computing  the dispersion of these values.
The errors grow as $p$ increases.
This is because moments of higher order are sensitive to the
tail of the distribution of the local edge variance. 
The  fit is 
such that the following average quadratic error:
$ E = \protect\sum_p 
\protect\frac {[\protect\rho(p,2)_{exp} - \protect\rho(p,2)_{th}]^2}
{b_p}$
is minimized.
We have checked that a gaussian dataset 
of images does exhibit ESS although it can not be explained by the 
SL model.
}
\label{rho.gr}
\end{center}
\end{figure}

First we show that SS holds in a range of scales $r$ with exponents
$\tau_{h,p}$ and $\tau_{v,p}$.
This is illustrated in Fig. (\ref{SS.gr}) where
the logarithm of the moments of the vertical and horizontal edge 
variances 
(as defined in eq. (\ref{eq:edge-horiz}) for the horizontal case)
are plotted as a function of $\ln r$.
Next we test ESS. The results are shown in Fig. (\ref{ESS.gr}) where
a linear behaviour of $\ln < \epsilon_{l,r}^{p} >$ vs 
$\ln < \epsilon_{l,r}^{2} >$ is observed 
in both the horizontal ($l=h$) and the vertical ($l=v$) directions.
One can see that ESS is valid
in a wider range than SS. This is similar to what is found in turbulence,
where this property has been used to obtain a more accurate
estimation of the exponents of the structure functions
(see e.g. \cite{all} and references therein).
The horizontal and vertical exponents
$\rho_h(p,2)$ and $\rho_v(p,2)$, estimated with
a least squares regression, are shown on Fig. (\ref{rho.gr}) as a 
function of $p$. From figs. (\ref{SS.gr}-\ref{rho.gr}) one sees that the 
horizontal and vertical directions 
have similar statistical properties,
which was not expected
(e.g. trees tend to increase luminosity correlations 
in the vertical direction). 
The SS exponents differ, as can be seen in Fig(\ref{SS.gr}).
What is even more surprising is that ESS not only holds for the 
statistics 
in both directions, but it does it with the {\em same} ESS exponents, 
i.e. $\rho_h(p,2) \sim \rho_v(p,2)$, within our numerical accuracy.

\begin{figure}[htb]
\hspace*{1.6cm}$P$
\begin{center}
\leavevmode
\epsfysize=10cm
\epsfbox[50 50 410 302]{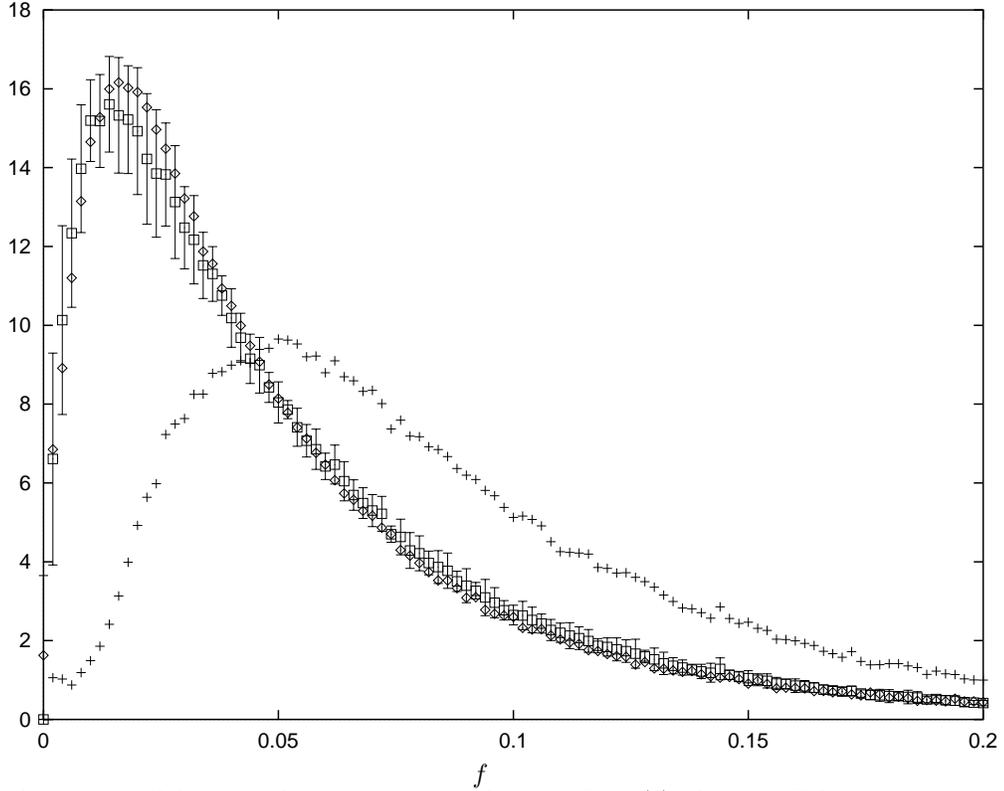}
\\
$f$
\\
\caption{ Verification of the validity of the integral representation of ESS, 
eq.(\protect\ref{G}) with a log-Poisson kernel,  for horizontal local edge variance. 
The largest scale is $L = 64$. Starting from the histogram 
$P_L(f)$ 
(crosses), and using a log-Poisson distribution with parameter 
$\beta = 0.50$ for the kernel 
$G_{rL}$, eq.(\protect\ref{G}) gives a prediction for the distribution at 
the scale $r=16$ (squares). This has to be compared with the 
direct evaluation of $P_r(f)$ (diamonds). 
Similar results hold for other pairs of scales. 
The error bars have been estimated as follows: 
The data set was divided in 9 groups as explained in the 
previous figure and the histograms at the scales $L$ and $r$ 
were computed for each group. Then for each group the histogram at scale $L$ 
was used to obtain a prediction for the histogram at scale $r$. 
The differences 
between the predicted and the computed values were squared and 
averaged over the groups. Its square root gives a measure of the error 
committed {\it in the prediction}, represented by the error bars.  
The test for vertical case is as good as for
horizontal variable.}
\label{histo.gr}
\end{center}
\end{figure}

Let us now consider scaling models 
to predict the $p$-dependence of the ESS exponents $\rho_l(p,2)$.
Since ESS holds, the SS exponents $\tau_{l,p}$ can be obtained from the 
$\rho_l(p,2)'s$ by measuring $\tau_{l,2}$.
The simplest scaling hypothesis is that, for a random variable $\epsilon_{r}({\bf x})$
observed at the scale $r$ (such as $\epsilon_{l,r}({\bf x})$), its probability distribution
$\bar{P}_r(\epsilon_{r}({\bf x})= \epsilon)$
can be obtained from any other scale $L$ by
\begin{equation}
\bar{P}_r(\epsilon) =\frac{1}{\alpha(r,L)} \; \bar{P}_L\left( \frac{\epsilon}{\alpha(r,L)} \right)
\label{naive}
\end{equation}
 From this one derives 
that $\alpha(r,L) = [\frac {<\epsilon_{r}^p>}{<\epsilon_{L}^p>} ]^{1/p}$ 
for any $p$, 
and 
that $\rho(p,2) \propto p$. If 
SS holds,
then  $\tau_p \propto p$:  
for turbulent flows 
this corresponds to the Kolmogorov prediction for the SS exponents \cite{K41}.
The nonlinear behaviour observed on Fig (\ref{rho.gr}) shows that this naive scaling
is violated 
(This is similar to what was observed in turbulence \cite{AGHA}
where the nonlinear behaviour was interpreted  
as evidence of the multifractal character of the turbulent flows \cite{PF85}).
This discrepancy becomes more dramatic if  
eq. (\ref{naive}) is expressed in terms of a normalized  
variable. Taking $\epsilon_r^{\infty} = \lim_{p \rightarrow \infty} 
< \epsilon_r^{p+1} >/ < \epsilon_r^{p} >$ the new variable is 
defined as $f_{r} = \epsilon_{r} / \epsilon_{r}^{\infty}$. 
If $P_r(f)$ is the distribution of $f_{r}$ the scaling 
relation, eq.(\ref{naive}), reads 
$P_r(f) = P_L(f)$. That this identity does not hold can be observed 
in Fig. (\ref{histo.gr}). 
A way to generalize this scaling hypothesis is to say that, 
instead of having one value
of $\alpha$ as in (\ref{naive}), every value of $\alpha$
contributes with a given weight. 
One then has: 
\begin{equation}
P_r(f) = \int G_{rL}(\ln \alpha) \frac{1}{\alpha}
P_L\left( \frac{f}{\alpha} \right) d\ln \alpha
\label{G}
\end{equation}
This scaling relation has been first introduced in the context
of turbulent flows \cite{CGH,SheLeveque,Dubrulle,Castaing}.
One can 
see that eq. (\ref{G}) is an 
integral representation of ESS with general 
(not necessarily linear) exponents. 
Once a 
kernel $G_{rL}$
is chosen the  $\rho(p,2)$'s can be predicted.

The difference between eqs. (\ref{naive}) and (\ref{G}) can 
also be phrased in terms of multiplicative processes 
\cite{Novikov,multiaffine}.  
Instead of $f_r \sim f_L$ we now have 
$f_r \sim \alpha f_L$ where 
the factor 
$\alpha$ 
itself becomes a stochastic 
variable determined by the kernel $G_{rL}(\ln \alpha)$. 
Since the scale $L$ is 
arbitrary (scale $r$ can be reached from any other scale $r'$) 
the kernel must obey a composition law. 
This stochastic variable at scale $r$ can then 
be obtained through a cascade of infinitesimal processes 
$G_{\delta} \equiv  G_{r,r + \delta r}$. 

Specific choices of $G_{\delta}$ define different models of ESS. 
The She-Leveque (SL) \cite{SheLeveque} model corresponds to  
a simple process such that $\alpha$ is 1 
with some probability $1-s$ and  is a constant   
$\beta$ with probability $s$. 
One can see that 
$s = \frac {1}{1-\beta^2} \ln (\frac {<f_{r+\delta r}^2>}{<f_{r}^2>})$  
and that this stochastic process 
yields a log-Poisson distribution for $\alpha$ \cite{SheWy}. 
It also gives ESS with 
exponents $\rho(p,q)$ that can be expressed in terms of a  
single parameter ($\beta$) as follows \cite{SheLeveque}:
\begin{equation}
\rho(p,q) = \frac{1-\beta^p - (1-\beta) p}{1-\beta^q - (1-\beta) q}
\label{SL}
\end{equation}

We have tested the model with the ESS exponents obtained 
with the image data set. 
The resulting fit for the SL model is shown in Fig. (\ref{rho.gr}). 
Both the vertical and horizontal ESS exponents can be  
fitted with  $\beta = 0.50 \pm 0.03$.
Other, more complex processes than log-Poisson, 
involving more than one parameter
have also been studied. 
We have also tested the model proposed in \cite{Castaing}. 
For our data, the best fit appears to be with the SL model, 
which is the simplest non trivial one.

The integral representation of ESS, eq. (\ref{G}), 
can also be directly tested on the 
probability distributions $P_r(f)$ and 
$P_L(f)$ evaluated from the data.  
In Fig. (\ref{histo.gr}) we show  the prediction for the 
distribution at the scale $r$ obtained from the distribution 
at the scale $L$. No new parameter is needed for this.

The parameter $\beta$ has allowed us to 
obtain all the ESS exponents $\rho(p,2)$. In order to 
obtain the SS exponents $\tau_p$ we need another 
parameter, e.g. $\tau_2$.
Let us first notice that,
for large $r$, 
$\epsilon_r^{\infty} \propto r^{\frac{\tau_2}{1-\beta}} 
\equiv r^{-\Delta}$. From the definition of $\epsilon_r^{\infty}$ one 
sees that it is controlled by the tail of the distribution 
$\overline{P}_r(\epsilon)$. This implies that the most singular structure
is the set of points where $\epsilon_r = \epsilon_r^{\infty}$.
Now a standard argument on multifractal scaling (see e.g. \cite{Frisch,PV87})
will relate the exponent $\Delta$ to the dimension $D_{\infty}$ of
this most singular structure. One finds:
$D_{\infty} = d - \frac{\Delta}{1-\beta}$
where $d=2$ is the dimensionality of the problem.
Since $\tau_p = \tau_2 \;\rho(p,2)$, a fit of  
$\tau_p$ determines $\Delta$. 
This was done for both the vertical and horizontal 
variables obtaining $\Delta_h = 0.4 \pm 0.2$ 
and $\Delta_v = 0.5 \pm 0.2 $ 
and leading to $D_{\infty,h}=1.3\pm0.3$ and $D_{\infty,v}=1.1 \pm0.3$. 
The quoted errors are purely statistical, but other 
sources of errors (e.g. the onset of the SS behavior) reduce the accuracy. 
As a result, we can say that
$D_{\infty,v} \sim D_{\infty,h} \sim 1$:
the most singular structures are almost one-dimensional, 
this reflects the fact that the most singular manifold 
consists of  sharp edges. 

To conclude we insist on the main result of this work, which is the existence of
non trivial scaling properties for the local edge variances. This  
property appears very similar to the one observed in turbulence 
for the local energy dissipation. 
In fact, we have seen that the SL model predicts all the relevant 
exponents and that, in particular, it describes the scaling behavior
of the sharpest edges in the image ensemble. 
A similar analysis could be performed taking into account color
or motion (analysing video sequences).
It would also be interesting to have a simple generative model of images which 
 - apart from having the correct power spectrum as in \cite{Ruderman-model} - 
would reproduce the self-similar properties found in this work.

\medskip 
\begin{center} 
{\bf Acknowledgements}
\end{center} 
We are grateful to Dan Ruderman for giving us his image data base.
We warmly thank Bernard Castaing for very stimulating discussions 
and Zhen-Su She for a discussion on the link between the scaling exponents
and the dimension of the most singular structure.
We thank Roland Baddeley and Patrick Tabeling for fruitful discussions.
We also acknowledge Nicolas Brunel for his collaboration during
the early stages of this work.
This work has been partly supported by the French-Spanish program "Picasso" and 
an E.U. grant.


\newpage

\onecolumn

\newpage

\newpage

\newpage

\end{document}